\documentclass[sigconf]{acmart}

\usepackage{listings}
\usepackage[many]{tcolorbox}  
\usepackage{tabularx}
\usepackage{svg}
\usepackage{amsmath}
\usepackage{multirow}
\usepackage{graphicx}
\usepackage{subcaption}
\usepackage{algorithm}
\usepackage{algpseudocode}
\usepackage[normalem]{ulem}
\newtcolorbox{boxA}{
    fontupper = \bf,
    boxrule = 1.5pt,
    colframe = black 
}
\newif\ifshowcomments
\showcommentsfalse

\newcommand{\jack}[1]{\ifshowcomments\textcolor{blue}{{\it [Jack says: #1]}}\fi}

\AtBeginDocument{%
  \providecommand\BibTeX{{%
    \normalfont B\kern-0.5em{\scshape i\kern-0.25em b}\kern-0.8em\TeX}}}

\begin{document}

\title{MemoCoder: Automated Function Synthesis using LLM-Supported Agents}

\author{Yiping Jia}
\email{yiping.jia@queensu.ca}
\affiliation{%
  \institution{Queen's University}
  \city{Kingston}
  \state{Ontario}
  \country{Canada}
}

\author{Zhen Ming Jiang}
\email{zmjiang@yorku.ca}
\affiliation{%
  \institution{York University}
  \city{Toronto}
  \state{Ontario}
  \country{Canada}
}

\author{Shayan Noei}
\email{s.noei@queensu.ca}
\affiliation{%
  \institution{Queen's University}
  \city{Kingston}
  \state{Ontario}
  \country{Canada}
}

\author{Ying Zou}
\email{ying.zou@queensu.ca}
\affiliation{%
  \institution{Queen's University}
  \city{Kingston}
  \state{Ontario}
  \country{Canada}
}

\begin{abstract}
With the widespread adoption of Large Language Models (LLMs) such as GitHub Copilot and ChatGPT, developers increasingly rely on AI-assisted tools to support code generation. While LLMs can generate syntactically correct solutions for well-structured programming tasks, they often struggle with challenges that require iterative debugging, error handling, or adaptation to diverse problem structures. Existing approaches such as fine-tuning\jack{You mentioned fine tuning here, but why you are not comparing against your approach with fine-tuning?} or self-repair strategies either require costly retraining or lack mechanisms to accumulate and reuse knowledge from previous attempts.

To address these limitations, we propose MemoCoder, a multi-agent framework that enables collaborative problem solving and persistent learning from past fixes. At the core of MemoCoder is a Fixing Knowledge Set, which stores successful repairs and supports retrieval for future tasks. A central \textbf{Mentor Agent} supervises the repair process by identifying recurring error patterns and refining high-level fixing strategies, providing a novel supervisory role that guides the self-repair loop. We evaluate MemoCoder across three public benchmarks---MBPP, HumanEval, and LiveCodeBench---spanning a range of problem complexities. Experimental results show that MemoCoder consistently outperforms both zero-shot prompting and a Self-Repair strategy, with improvements ranging from 3.1\% to 12.1\% in Pass@10 and from 1.4\% to 14.5\% in Pass@50, demonstrating its effectiveness in iterative refinement and knowledge-guided code generation.

\end{abstract}

\maketitle

\section{Introduction}

With the introduction of Large Language Models (LLMs), software development has witnessed significant advancements. LLMs have demonstrated remarkable capabilities in performing various coding tasks such as code completion, bug fixing, and code generation~\cite{wang2023codet5+,pan2025codecor,nijkamp2022codegen}. LLM-based coding assistants, such as GitHub Copilot\footnote{https://copilot.microsoft.com} and ChatGPT\footnote{https://chatgpt.com/}, have gained widespread adoption, with surveys indicating that over 50\% of developers now use AI-assisted tools in their development process\footnote{https://survey.stackoverflow.co/2024/}. Studies have also shown that developers using these tools complete tasks significantly faster than those relying solely on traditional methods~\cite{dohmke2023seachangesoftwaredevelopment,peng2023impactaideveloperproductivity}.

However, LLMs face substantial challenges when handling programming tasks~\cite{austin2021programsynthesislargelanguage}. For example, LLMs can often generate syntactically correct and logically coherent code for routine tasks with clear specifications. In addition, their performance degrades on problems requiring multi-step reasoning, algorithm design, or generalization across diverse input-output structures~\cite{naveed2024comprehensiveoverviewlargelanguage,austin2021programsynthesislargelanguage}. Furthermore, errors in the generated code often require manual debugging, which limits the productivity benefits of these models~\cite{dou2024whats,kaniewski2024vulnerabilityhandlingaigeneratedcode}. These challenges motivate the need for approaches that enhance the reliability, adaptability, and effectiveness of LLM-based code generation.

Recent studies have explored fine-tuning LLMs for code generation~\cite{finetuningLLMforsecureCodeGen,weyssow2024exploringparameterefficientfinetuningtechniques,yuan2023evaluatinginstructiontunedlargelanguage} and leveraging fixing-and-looping strategies where an LLM iteratively refines its output during inference~\cite{jiang2023selfevolvecodeevolutionframework,olausson2024selfrepairsilverbulletcode}. Fine-tuning requires modifying model weights using supervised training, which is resource-intensive and inflexible as it must be retrained whenever new examples or error types are introduced~\cite{moller2024prompt}. Fixing-and-looping\jack{cite?}, by contrast, keeps the model frozen and instead relies on repeated prompting to guide corrections. However, such methods lack memory across tasks and iterations, forcing the model to re-discover similar fixes repeatedly. These approaches leads to inefficiency, limited generalization, and no accumulation of reusable knowledge.

To address these challenges, we propose MemoCoder, a multi-agent code generation framework that enables collaborative problem-solving and persistent learning from past fixes. MemoCoder consists of four specialized LLM-based agents (Planner, Code Writer, Test Executor, and Mentor) and a memory module. The \textit{code writer} generates initial code from a problem description. The \textit{code repairer} iteratively fixes errors using retrieved suggestions. The \textit{mentor} distills high-level error patterns and refines repair strategies over time. The memory module, which stores successful repair examples that agents repair the code generation errors and supports retrieval during subsequent tasks. 
These agents interact to form an adaptive and reusable repair loop that supports robust code generation. The key contribution of our approach is two-fold: (1) the Mentor Agent, which plays a supervisory role in the system, monitors the types of errors encountered during the repair, summarizes recurring error patterns, and dynamically updates fixing strategies based on newly acquired insights by retrieving relevant information from the fixing knowledge set. (2) The memory module enables MemoCoder to continually learn new and apply existing repair strategies without retraining. Together, the Memory Module and the Mentor Agent allow MemoCoder to generalize learned repairs across different programming problems described by natural language, reducing redundant exploration and improving repair efficiency.\jack{Either do Mentor Agent, or mentor agent, for case consistency. fix throughout}

We evaluate MemoCoder using three public benchmarks—LiveCodeBench (LCB)~\cite{lcb}, Mostly Basic Python Problems (MBPP)~\cite{MBPP}, and HumanEval (HE)~\cite{HumanEval}—which span a range of task complexities and test the system's ability to iteratively refine code and generalize learned repairs. Our method is compared against two strong baselines: zero-shot prompting and a Self-Repair strategy. Experimental results show that MemoCoder consistently outperforms both baselines, with improvements ranging from 3.1\% to 12.1\% in Pass@10 and from 1.4\% to 14.5\% in Pass@50, demonstrating its effectiveness in iterative refinement and knowledge-guided code generation.

\textbf{Our contributions are summarized as follows:}
\begin{itemize}
    \item We propose MemoCoder, a multi-agent framework that enables role-specialized collaboration among LLMs for code generation and repair, improving the test pass rate for the generated code through planning algorithmic strategies, generating and refining code, executing tests, and providing feedback based on prior fixes.\jack{Present some results w.r.t. to baseline highlight its value}
    \item We introduce a persistent \textit{Fixing Knowledge Set} that accumulates successful repairs across tasks and supports retrieval to guide future error correction.\jack{I would say the framework is flexible and adaptive as they can continuously learn new knowledge without rediscovering bug fixing strategies for recurrent problems}
    \item We design a novel \textit{Mentor Agent} that analyzes common error patterns across failed attempts and distills reusable fixing suggestions, allowing the system to continually improve repair quality across tasks. \jack{I would say the framework is flexible and adaptive as they can continuously learn new knowledge without rediscovering bug fixing strategies for recurrent problems. Say case studies show that xx\% of problems are recurrent problems or can be effectively addressed with the existing knowledge in xxx, which can be successfully fixed by our approaches instead of rediscovering the diagnosis and fixing strategies}
\end{itemize}

The remainder of this paper is structured as follows. Section~\ref{sec:experiment_setup} introduces MemoCoder in detail, including the design and interaction of the agents. Section~\ref{sec:results} presents our experimental questions, along with the corresponding evaluation process, results, and analysis. Section~\ref{sec:threat} discusses the threats to validity. Finally, Section~\ref{sec:conclusion} concludes the paper and outlines directions for future work.

\section{Overview of MemoCoder}
\label{sec:experiment_setup}

\begin{figure*}
  \centering
  \includegraphics[width=\linewidth]{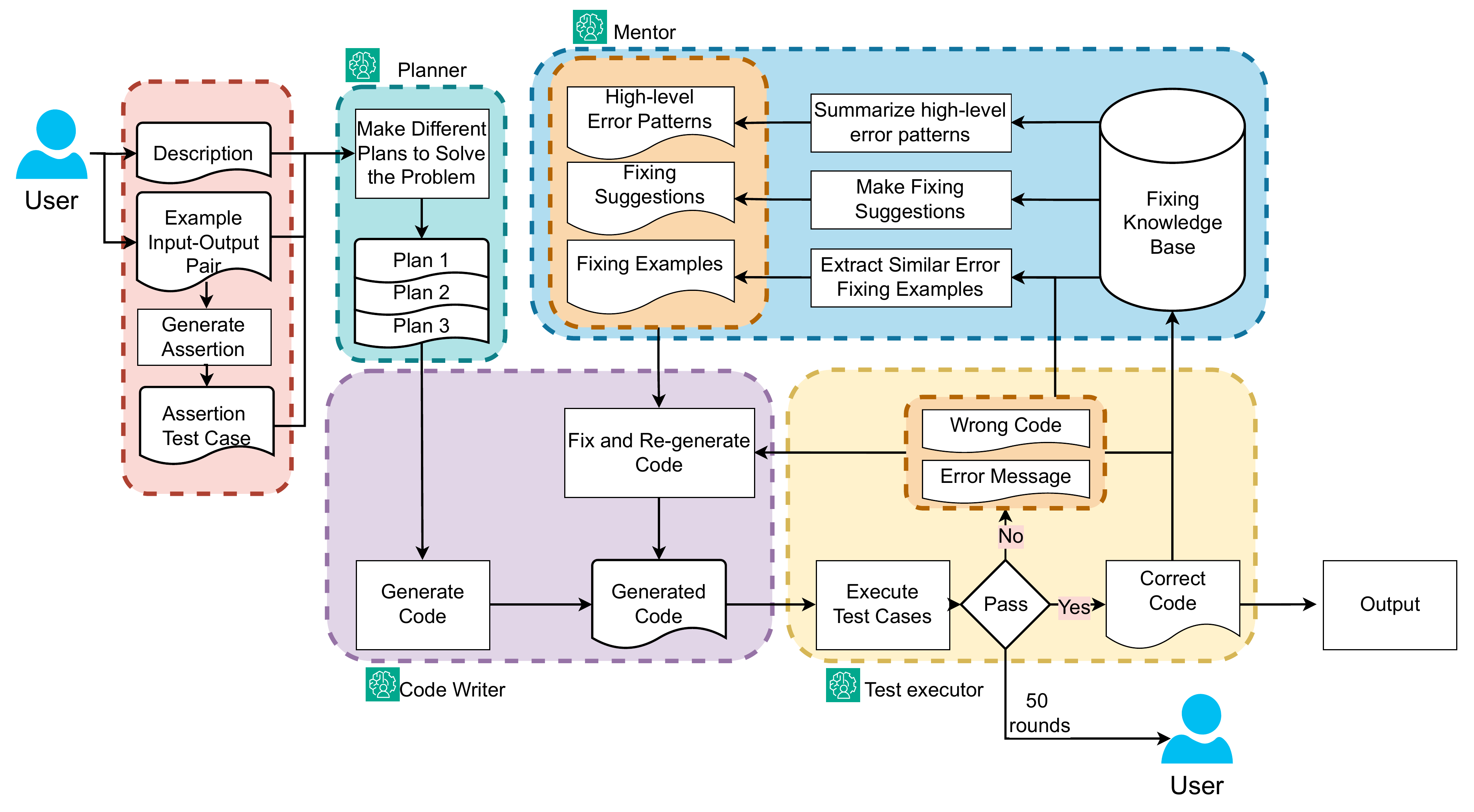}
  \caption{An overview of MemoCoder.}
  \label{img:approach}
\end{figure*}

\jack{Before this section, can you have motivation section? IN there I would list some problems that you would like to tackle: One example when directly promtp the LLMs, it generates hallucinated results.  1 - 2 more examples, showing some of the errors (prefer one syntaical and one semamtical) are common across various problems/runs. Then you motivate this by saying our approach directly tackles these problems}

An overview of MemoCoder is illustrated in Figure~\ref{img:approach}. MemoCoder consists of four LLM-based agents, namely \textit{Planner}, \textit{Code Writer}, \textit{Test Executor}, and \textit{Mentor}, which work collaboratively to solve programming problems.\jack{These do not match with the intro component.}
The Planner generates multiple algorithmic strategies for solving a given problem. The Code Writer implements one selected plan as an initial solution by generating a method-level code. The Test Executor runs the generated code against a guiding test case; if it passes, the solution is provided as the output from MemoCoder. Otherwise, the error is passed to the Mentor, which retrieves relevant past fixes and provides a fixing suggestion. The Code Writer then uses this feedback to iteratively revise the code. This process continues until a correct solution is found or a retry limit is reached. Once the generated code is fixed, the fixed code and the initial generated code are restored in a fixing knowledge set.
\jack{The approach description is a bit confusing to follow. I would suggest the following: 

Add number in the diagram and some short descriptions for steps. For example, (1) Task Planning, (2) Initial Code Generation, (3) Testing and Verification, (4) Error Diagnosis and Fix Suggestions, (5) Code Revision. 

Describe your running example. 

For each step, you first describe why you need to do that (e.g., without planning), the LLM may hallucinate. Then how you do that with the running example.  

I am not sure why this different phases (training vs. applying), as we are always learning and applying knowledge. If yes, please clearly state here. 
}

\subsection{Test Case Generation} 
\label{testcasegeneration}
It is essential to test the generated code using test cases. 
In particular, the test cases enable us to systematically assess the functional correctness of the generated code and benchmark the performance of MemoCoder under controlled and reproducible settings. 
In our work, we aim to generate test cases to include the semantic information such as the expected function name, the structure and types of input arguments, and the format of the output.
A test case also provides a concrete example of how the function should behave, which helps the model better understand the intended functionality.
Test cases can be automatically executed after each time the code is generated, allowing us to immediately check whether the solution behaves correctly. 
Test cases are implemented as the JUnit style assertions. If an assertion fails, it raises an exception, indicating a mismatch between the expected and actual results, which helps identify bugs during testing. Test cases provide execution feedback, which is especially useful for guiding the \textit{code writer} agent to target fixes based on actual runtime errors or failed conditions.\jack{You still did not explain how you generate test cases here. Are they extracted from the original problem descriptions or automatically derived by LLMs? Your running examples should start here}

For each problem, we ensure that we have at least three assertion test cases. Among them, the first assertion is designated as the \textbf{guiding assertion}, which is used throughout the code generation and repair process to direct the agents toward producing a correct solution for a specific example. This allows the system to focus on achieving functional correctness in a controlled setting. The remaining tests stay hidden until the evaluation phase. This procedure ensures that the generated code does not overfit to the provided test cases and can still pass unseen ones, indicating that the model has learned the underlying logic rather than memorizing the visible cases.

Figure~\ref{fig:example_problem} shows an example problem along with the expected input-output pairs with out constructed assertion test cases. These assertions ensure that the function \texttt{min\_cost} correctly computes the minimum cost path and returns the expected output for different testing scenarios. \jack{In here what's the guiding assertion then? Or all three of them are?}

\begin{figure}
\begin{tcolorbox}[
    colback=gray!5,
    colframe=black,
    sharp corners=south,
    boxrule=0.5pt,
    enhanced,
    before skip=10pt,
    after skip=10pt
]
\scriptsize
\textbf{Task:} \\
Write a function to find the minimum cost path to reach (m, n) from (0, 0) for the given cost matrix cost[][] and a position (m, n) in cost[][].

\textbf{Input-Output Pairs:}
\begin{itemize}
    \item \texttt{Input: ([[1, 2, 3], [4, 8, 2], [1, 5, 3]], 2, 2)} \\
          \texttt{Output: 8}
    \item \texttt{Input: ([[2, 3, 4], [5, 9, 3], [2, 6, 4]], 2, 2)} \\
          \texttt{Output: 12}
    \item \texttt{Input: ([[3, 4, 5], [6, 10, 4], [3, 7, 5]], 2, 2)} \\
          \texttt{Output: 16}
\end{itemize}
\textbf{Constructed assertion cases:}
\begin{itemize}
\item\texttt{assert min\_cost([[1, 2, 3], [4, 8, 2], [1, 5, 3]], 2, 2) == 8}
\item\texttt{assert min\_cost([[2, 3, 4], [5, 9, 3], [2, 6, 4]], 2, 2) == 12}
\item\texttt{assert min\_cost([[3, 4, 5], [6, 10, 4], [3, 7, 5]], 2, 2) ==16}
\end{itemize}

\end{tcolorbox}
\caption{An example problem description with an input-output pair}
\label{fig:example_problem}
\end{figure}





\subsection{Agent Design}
\label{subsec:agent_design}

MemoCoder decomposes the task of program synthesis and repair into four specialized roles: \textit{Planner}, \textit{Code Writer}, \textit{Test Executor}, and \textit{Mentor}. Each role is fulfilled by an LLM-based agent that operates based on structured prompt templates. These agents interact to iteratively refine the generated code until a correct solution is found or the maximum number of attempts is reached.

\subsubsection{Planner}
During the training phase\jack{You first mention this. You need to explain the phases}, we observe that directly asking an LLM to write the code for the function sometimes leads to hallucinated or off-track implementations that provide a poor starting point for later repair~\cite{agarwal2024codemiragehallucinationscodegenerated}. Therefore, we implement a Planner agent as the first agent that aims to make multiple plans before any code is generated to help the LLM anchor the reasoning of the model~\cite{zhang2023planninglargelanguagemodels,jiang2024selfplanningcodegenerationlarge}, reduce hallucination, and increase the chance that at least one plan is close to the correct algorithm.

The \textit{Planner} agent initiates the problem-solving process by producing diverse algorithmic strategies for solving the problem. Given the natural language description and the guiding assertion, the Planner generates three distinct high-level plans. Each plan is a step-by-step natural language explanation that outlines the logic and algorithmic techniques used to solve the problem. These plans help the system explore the solution space from multiple perspectives. The generated plans are then passed to the \textit{Code Writer} for implementation.

\subsubsection{Code Writer}\jack{Maybe just the Coder Agent, as you have the Planner Agent, The Tester Agent, the Mentor Agent?}

The \textit{Code Writer} serves two key roles: (1) selecting a plan and generating the initial implementation, and (2) iteratively refining faulty code during the repair loop. Initially, the Code Writer selects one of the plans proposed by the Planner and attempts to implement a function that satisfies the provided guiding assertion. The input to the Code Writer includes the problem description, the implementation plans generated by the Planner, and the guiding assertion test case.\jack{Did you generate many candidate solutions at one time? If not, how did you calculate pass@k values?}

The Code Writer is also responsible for fixing the faulty\jack{faulty or buggy? pick one throughout} code if the code does not pass the guiding test case. If the initial implementation fails, the Code Writer enters the repair loop and tries to re-generate the code based on:
\begin{itemize}
\item \textit{The original problem description}, which ensures that the \textit{code writer} understands the intended functionality of the task, offering essential context about the expected behaviour of the code.

\item \textit{The initial code snippet produced by the Code Writer\jack{The Coder Agent}}, which  provides the starting point for debugging. Analyzing the original implementation helps the code writer identify where and how the logic may have deviated from the intended behaviour.

\item \textit{The guiding assertion}, which serves as a concrete test case that the fixed code must pass. It defines the expected behaviour and acts as an objective reference for validating the correctness of any revision.

\item \textit{The error type of the execution} helps the code writer narrow down the class of issue encountered (\textit{e.g.}, syntax error, runtime exception, logical error), allowing it to apply more targeted fixing strategies.

\item \textit{The detailed error message} provides fine-grained diagnostic feedback from the Python interpreter or runtime environment. This message often pinpoints the location or cause of the error and plays a crucial role in guiding effective code modifications.

\item \textit{A fixing suggestion} offers high-level guidance on how to address the identified error type. This information is retrieved from the \textit{mentor} agent which is explained in Section~\ref{sec:results}
\end{itemize}
Using this feedback, the Code Writer attempts to revise the code iteratively until it satisfies the test case or exceeds the maximum repair limit.\jack{Running example?}

\subsubsection{Test Executor}\jack{The Tester Agent}

The \textit{Test Executor} runs the code produced by the Code Writer and evaluates it against the guiding assertion using a deterministic script-based procedure. If the generated code passes the assertion test, it is considered correct and will be processed for the user. When the code fails the assertion test, it is automatically categorized into four different error types based on the code execution: 
\begin{itemize}
\item \textit{Compile Error:} The code fails to compile due to issues such as invalid syntax or missing imports. 
\item \textit{Exceptions:} The code compiles but runs into errors during the execution, and it fails to give any output. 
\item \textit{Assertion Failure} The code compiles but yields incorrect results for the test. 
\item \textit{Timeout Errors:} The code does not terminate within a set time limit (5 seconds in our experiments), often due to infinite loops or similar runtime issues. 
\end{itemize}

If the code fails to pass a test case, the \textit{test executor agent} records the error type and the corresponding messages and passes the error type, the error message and the code to fix to the \textit{mentor agent}.\jack{Running example?}

\subsubsection{Mentor Agent} 
The Mentor agent plays a supervisory role in MemoCoder, overseeing the code repair process by the code writer agent and ensuring that fixing strategies get updated over time. To have enough starting point for knowledge retrieval, we start retrieving when at least 20 new successful fixes for a given error type are added during the knowledge accumulation phase (explained in Section~\ref{subsec:data_preparation}). The Mentor (1) analyses the new examples, (2) summarizes the high-level causes of the given error type, and (3) rewrites the short list of “fixing suggestions’’ for that error type. Unlike the code writer, which focuses on resolving individual errors, the Mentor analyzes patterns across multiple repair attempts to refine and optimize the system’s fixing strategies by summarizing high-level error patterns and providing suggestions on fixing the error type.

The Mentor's responsibility is to maintain and dynamically update fixing suggestions for each type of error. For each error type, a predefined fixing suggestion summarizes the common issues associated with that error and outlines possible solutions. The suggestions serve as reference points for the \textit{code writer} agent, helping it apply targeted fixes more efficiently later.

During evaluation, we adopt an \textit{online‐adaptation} protocol~\cite {tack2024onlineadaptationlanguagemodels}:
All test problems are processed iteratively in successive rounds. In each round, only the remaining unsolved problems are passed through the system, which incorporates newly distilled fixing strategies from previously solved cases. This online adaptation ensures causal ordering and prevents data leakage, as no task benefits from its own fixes. \jack{Running example?}

All prompts for the agents are also included in the replication package~\footnote{https://anonymous.4open.science/r/memoCoder-3BD2} for transparency.

\section{Experiment setup}

\subsection{Knowledge Data Curation}
~\label{subsec:data_preparation}
To form the dataset for our study, we use coding competition datasets, which provide well-defined programming problems commonly used to evaluate code generation models. These problems typically require writing a single function that takes specific inputs and returns the correct output as defined by the problem statement. These four selected datasets include: (1) \textit{APPS}~\cite{APPS}, containing 10,000 problems; (2) \textit{LCB}~\cite{lcb}, which includes 500 problems; (3)~\textit{HE}~\cite{HumanEval}, with 164 problems; and (4) \textit{MBPP}~\cite{MBPP}, comprising 974 problems. These datasets provide a diverse range of programming problems gathered from different resources, such as basic logical programming, interview-level programming questions and contest-level challenges, making them suitable for both knowledge accumulation and evaluation purposes.
MBPP and HumanEval are widely used in the code generation literature and serve as standard benchmarks for evaluating functional correctness. They consist of short Python programming problems with accompanying unit tests, allowing for easy comparison with prior work. LCB contains more diverse and realistic programming tasks, many of which are drawn from competitive programming settings. It also supports contamination-controlled evaluation by allowing us to filter problems based on model pretraining cut-off dates, providing a more rigorous test of generalization.

\jack{I assume none of the datasets have readily available test cases. how you extract test cases should go to your approach section instead. }The four datasets used in our study differ in the way problem descriptions are structured. For example, \textbf{MBPP} supplies a ready-made list of assertion-style test cases including the required function name and explicit input–output pairs, whereas \textbf{APPS} provides only the function name and raw input–output examples, leaving test-case construction to the user. To ensure consistency and fair comparison across datasets, we pre-process the questions to include the following attributes for each problem:
\begin{itemize}
    \item  \textit{Description of the problem}: The textual explanation of the problem to be solved. All four datasets already provide a natural–language problem statement, which we adopt in their original form.
    \item  \textit{Name of the function}: The expected name of the function to be generated. The required function name is taken directly when the dataset lists it (as in \textit{APPS}); if the name appears only inside ready-made assertions (as in \textit{MBPP}), we read it from the first assertion.
    \item \textit{Test cases}: The test cases we use to assert the function outputs. When the dataset already gives full assertions, we keep them. Otherwise, we wrap each input–output pair in the form \verb|assert f(x) == y|.  Assertions are commonly used to verify expected behaviour by checking whether a specific condition holds.
\end{itemize}

In our proposed method for solving programming problems, we use knowledge from previously solved programming problems to build a knowledge base to address future problems. More specifically, to bootstrap our approach, we apply our proposed approach to the largest dataset, \textit{APPS dataset}~\cite{APPS}, to form the foundation of our code fixes knowledge base (\textit{i.e.} training dataset). 
Three smaller datasets \textit{LCB}~\cite{lcb}, \textit{HumanEval}~\cite{HumanEval}, and \textit{MBPP}~\cite{MBPP} are then used to test our proposed approach, which utilizes previous code fixes obtained in the knowledge accumulation phase. We test our method on MBPP, HumanEval and LCB to ensure that these datasets remain unseen during the knowledge accumulation phase to avoid data contamination. However, we acknowledge that \textit{HumanEval} and \textit{MBPP} were publicly released well before the training cut-off dates of many large language models. As a result, even though these datasets are not included in our own training or knowledge base, models used in MemoCoder (e.g., \textit{LLaMA 3.1-8B-Instruct}, \textit{Qwen 2.5-32B}) may have indirectly seen similar or identical examples during pretraining. This presents a potential threat of contamination.

In contrast, \textit{LCB} is a dynamic benchmark that tracks the release dates of its problems. To minimize contamination risk, we follow LCB's guidelines and evaluate on a 6-month subset of problems that are explicitly labelled as free from training data contamination for the two models we use (i.e., LLaMA 3.1-8B-Instruct and Qwen 2.5-32B) according to the official leaderboard. This approach avoids manual verification of cutoff dates while ensuring a clearer separation between training and evaluation, providing a more reliable test of MemoCoder's generalization.



\subsection{Selection of Base Models}

In MemoCoder, we utilize four agents to perform code generation tasks: the Planner, Code Writer, Test Executor, and Mentor. Each agent, except Test Executor, is implemented using an LLM that can generate and refine code while following structured problem-solving processes. We structure and evaluate our approach using two different instruction-tuned models \textit{LLaMA 3.1-8B-Instruct} and \textit{Qwen 2.5-32B}, each serving as the foundational model for all agents within a given workflow. We select these two models for the following reasons: 
\begin{itemize}
    \item Both models are general-purpose and have been trained to handle both natural and programming languages, which is a critical requirement when developing autonomous coding agents. The ability to process problem descriptions, error messages, and code snippets ensures that agents can perform well in an interactive programming environment. 
    \item Using models of different sizes provides a more robust evaluation and comparison. By including both a small-sized model (LLaMA 3.1-8B) and a mid-sized model (Qwen 2.5-32B), we can better understand how model scale affects agent performance in code generation, refinement, and reasoning. 
    \item Both models are instruction-tuned, meaning they are optimized to follow human instructions and structured prompts effectively. This property is particularly important for developing agents, as they need to engage in an iterative problem-solving process that involves interpreting prompts, executing guided refinements, and dynamically adjusting code generation based on feedback.
    \item Both models demonstrate strong performance on the LCB leaderboard~\cite{lcb}, which evaluates LLMs on problems released over time. LLaMA 3.1-8B and Qwen 2.5-32B are among the top-performing models on the subset of problems released between April 1st and October 1st, 2024.
    \item The LCB leaderboard flags potential data contamination by comparing each problem's release date with the training cut-off dates of LLMs. We use problems with a six-month span released between April 1st and October 1st, 2024. Our selected models, LLaMA 3.1-8B and Qwen 2.5-32B, are confirmed to be contamination-free on this subset, unlike models such as QwQ and DeepSeek, ensuring the reliability and validity of our evaluation.

\end{itemize}

\textit{LLaMA 3.1}~\cite{llama3} is part of the Meta AI LLaMA series, known for its high performance across a wide range of natural language understanding and reasoning tasks. The 8B variant of LLaMA 3.1 is a compact yet capable model that balances efficiency and effectiveness, making it well-suited for tasks requiring both general reasoning and domain-specific knowledge. It has been optimized for both natural language processing and code-related tasks, making it a strong candidate for developing intelligent coding agents. 

\textit{Qwen 2.5}~\cite{qwen} is a larger, 32B-parameter model developed by Alibaba that has demonstrated strong performance across multiple benchmarks, including programming-related tasks. Compared to LLaMA 3.1-8B, \textit{Qwen 2.5-32B} has a significantly larger parameter count, which allows it to model more complex patterns in both natural and programming languages. This makes it particularly useful for handling intricate code generation and repair tasks that require deeper reasoning. 

We use the default temperature setting of 0.7 for both LLMs in our experiments\jack{as in xxx papers?}. This moderate temperature encourages diversity in generations without introducing excessive randomness. A lower temperature (e.g., 0.0) may cause the model to produce repetitive or deterministic outputs across iterations, which could reduce the effectiveness of the multi-agent repair process.

\subsection{Evaluation of Our Approach}
To assess the effectiveness of our proposed approach, we use the \textit{Pass@k} metric, a widely adopted evaluation measure for code generation models~\cite{lcb,zhang2023planninglargelanguagemodels}. The \textit{Pass@k} metric represents the percentage where at least one of the top $k$ generated solutions passes all testing assertions for given programming problems. \textit{Pass@k} evaluation reduces the impact of randomness in model outputs during evaluation by averaging over multiple outputs~\cite{chen2021evaluatinglargelanguagemodels}.
For example, consider a model that generates an initial solution which fails the test cases—this means \textit{Pass@1} is 0 (or False). If the model attempts to fix the code twice but the resulting solutions still fail, then \textit{Pass@3} is also 0. If the model finally generates a correct solution on the fourth try (i.e., the third fix after the initial output), then \textit{Pass@4} becomes 1 (or True), since a correct solution was found within the first four attempts.

This metric is useful for evaluating the model’s robustness across multiple trials. For instance, \textit{Pass@5} measures whether any of the five generated solutions is correct, reflecting the model's ability to eventually produce a valid solution within five tries.

We separate our proposed approach in two phases:
\begin{itemize}
    \item 
    \textit{Knowledge Accumulation Phase:} The MemoCoder is run on the APPS dataset to update fixing strategies and populate the program repair knowledge set. This phase prepares the knowledge base for downstream retrieval during code repair.\jack{I am not clear where do you use all three LLMs to do this step?}
\item 
    \textit{Evaluation Phase:} We evaluate our approach on three benchmark datasets—\textit{LCB}, \textit{MBPP}, and \textit{HumanEval}—using the knowledge accumulated from APPS.
    During this phase, the Fixing Knowledge set continues to grow by incorporating successful repairs, which are made available for retrieval in future test cases. 
\end{itemize}

By combining the two base models (\textit{LLaMA 3.1-8B-Instruct} and \textit{Qwen 2.5-32B}) with the three evaluation datasets (\textit{LCB}, \textit{MBPP}, and \textit{HumanEval}), we obtain six distinct experimental setups. These setups allow us to systematically evaluate the performance of our approach across different model sizes and problem types.

\jack{You need to motivate and bring up the RQs here before you talk about that in the next section}

\section{Results}
\label{sec:results}

This section presents motivation, approach, and findings for each of our research questions.

\subsection{RQ1: How effective is MemoCoder?}

\subsubsection{Motivation}
Previous research has demonstrated that code generation using large language models can result in problematic code, hallucinations, and unintended behaviors~\cite{agarwal2024codemiragehallucinationscodegenerated}. 
In this research question, we aim to evaluate MemoCoder against a one-shot LLM prompt and self-repair~\cite{olausson2024selfrepairsilverbulletcode,lcb} looping mechanisms, where the LLM attempts to fix its errors without employing distinct agents. We build on the self-repair implementation provided in the GitHub repository of LCB~\footnote{https://github.com/LiveCodeBench/LiveCodeBench}.

\subsubsection{Approach}

Our experiment consists of two phases: (1) a knowledge accumulation phase and (2) the evaluation phase (explained in Section~\ref{subsec:data_preparation}). 
We first construct the knowledge base (\textit{i.e.,} knowledge accumulation phase) using the APPS dataset. Code fixes serve as the basis for Retrieval-Augmented Generation (RAG), where the \textit{mentor agent} retrieves relevant examples and passes them to the \textit{code writer agent} to help the model learn from similar errors and fixed that have happened in the past. RAG is a hybrid framework that combines retrieval with LLMs. To determine the optimal number of examples to retrieve, we manually experimented with different retrieval sizes of retrieved examples,
and we find that retrieving ten examples offers the best trade-off between informativeness and attention. Larger retrieval sizes often led to truncation or loss of focus, as supported by recent findings on LLMs’ context length limitations~\cite{lu2024insightsllmlongcontextfailures,liu2023lostmiddlelanguagemodels}. Therefore, we set the mentor agent to retrieve up to ten examples from our knowledge base based on error similarity.\jack{These should go to yMemoCoder description section}

The knowledge retrieval in our RAG system within the mentor agent is based on error message similarity using a Longest Sequential Matching heuristic, which identifies the longest contiguous sequence of matching words between two error messages, considering only alphanumeric words and removing all punctuation. The rationale is that similar error messages often share common phrases or structures, and identifying the shared sequences can help retrieve relevant past fixes. For example:
\begin{itemize}
    \item Message 1: \textit{"not enough values to unpack (expected 2, got 1)"}
    \item Message 2: \textit{"too many values to unpack (expected 2)"}
\end{itemize}
The longest sequential match between these messages is five tokens (i.e., "values to unpack expected 2"), indicating a high likelihood of error similarity and fix applicability.\jack{These sould go to yMemoCoder description section}

We define a problem as fixed only if it passes all the defined assertion test cases for that problem. We evaluate MemoCoder using input problems from our evaluation datasets, namely \textit{LCB}, \textit{MBPP}, and \textit{HE}. We compare the performance of MemoCoder against two baseline methods:
\begin{itemize}
    \item \textbf{Zero-shot:} The base model generates code directly from the problem description without iterative refinement.\jack{Add citation saying as in xxx these papers}
    \item \textbf{Self-Repair:} The base model iteratively refines its own output using self-generated feedback, without knowledge retrieval via mentoring.\jack{Add citation saying as in xxx these papers}
\end{itemize}

We use the \emph{McNemar’s test}~\cite{McNemar_1947} to assess the statistical significance of differences in pass rates between MemoCoder and the baselines. McNemar’s test is specifically designed for \textit{paired binary} outcomes, which fits our setting where each code generation method is evaluated on the \textit{same set of problems}, and each problem yields a binary result (\textit{pass} or \textit{fail}). This test accounts for the paired nature of the data by focusing on the disagreement cases, i.e., the number of problems solved by one method but not the other, making it more appropriate than unpaired tests or comparisons of aggregate pass rates. It enables us to determine whether the observed improvements are statistically significant beyond random fluctuations in individual test cases.

The overall performance of MemoCoder compared to the baseline methods across the three benchmark datasets is summarized in Table~\ref{tab:rq1_results}.\jack{I don't get how you get your pass@k data, does yMemoCoder generate multiple candidate each time? if yes, how do they provide fixing suggestions for which results?}

\begin{table*}[t]
\centering
\caption{Passing rates of MemoCoder and baselines across LCB, MBPP, and HE for LLaMA 3.1-8B-Instruct and Qwen 2.5-32B. Results marked with a star ($^*$) indicate that the baseline performance is significantly lower\jack{not lower, different. McNemar can only tell if they are different} than MemoCoder, as determined by McNemar's test ($p < 0.05$).}
\label{tab:rq1_results}
\begin{tabular}{l|lllll|lllll}
\hline
\textbf{Dataset} & \multicolumn{5}{c|}{\textbf{LLaMA 3.1-8B-Instruct}} & \multicolumn{5}{c}{\textbf{Qwen 2.5-32B}} \\
& \textbf{Method} & \textbf{Pass@1} & \textbf{Pass@5} & \textbf{Pass@10} & \textbf{Pass@50} & \textbf{Method} & \textbf{Pass@1} & \textbf{Pass@5} & \textbf{Pass10} & \textbf{Pass@50} \\
\hline
\multirow{3}{*}{LCB} 
& Zero-Shot       & 15.62\% * & 29.57\%   & 34.76\%*  & 46.58\%*  & Zero-Shot       & 32.70\%*      & 50.21\%  &55.38\%*   & 63.71\%*\\
& Self-Repair   & 18.23\%   &28.40\%    & 31.47\%*  & 44.67\%*  & Self-Repair   & 36.28\%       & 47.60\%  &50.10\%*   & 59.93\%*\\
& MemoCoder          & 17.94\%   &32.62\%    & 45.52\%   & 51.06\%   & MemoCoder          & 35.13\%       & 54.97\%  &63.59\%    & 74.46\%\\
\hline
\multirow{3}{*}{MBPP} 
& Zero-Shot       & 54.68\%*  & 70.05\%   &73.06\%*   & 76.39\% *& Zero-Shot & 75.333\% *& 82.53\%&85.34\%* & 90.12\%*\\
& Self-Repair   & 59.67\%   &68.03\%    &70.22\%*   & 73.41\%*& Self-Repair & 79.21\% &82.09\% &84.69\%*& 87.07\%*\\
& MemoCoder          & 60.36\%   & 72.85\%   &77.69\%    & 83.23\% & MemoCoder & 79.56\%& 84.38\%& 89.72\%& 93.70\%\\
\hline
\multirow{3}{*}{HE} 
& Zero-Shot & 31.60\%& 40.61\% &43.38\%*& 50.97\%& Zero-Shot & 80.23\%& 86.20\%& 88.37\%*& 92.24\%*\\
& Self-Repair & 34.92\%& 39.99\%&42.60\%* &47.49\%*  & Self-Repair & 83.81\%&86.56\% &89.13\%*& 93.30\% \\
& MemoCoder& 33.49\%&43.34\% &48.91\%& 52.40\% & MemoCoder& 82.97\%& 88.19\%&92.21\% & 95.33\%\\
\hline
\end{tabular}
\end{table*}

\begin{table*}[t]
\centering
\caption{Ablation study: Passing rates of MemoCoder with different components removed across LCB, MBPP, and HE. for LLaMA 3.1-8B-Instruct and Qwen 2.5-32B.}
\label{tab:rq2_ablation}
\begin{tabular}{l|llll|llll}
\hline
\textbf{Dataset} & \multicolumn{4}{c|}{\textbf{LLaMA 3.1-8B-Instruct}} & \multicolumn{4}{c}{\textbf{Qwen 2.5-32B}} \\
& \textbf{Method} & \textbf{Pass@1} & \textbf{Pass@10} & \textbf{Pass@50} & \textbf{Method} & \textbf{Pass@1} & \textbf{Pass10} & \textbf{Pass@50} \\
\hline
\multirow{3}{*}{LCB} 
& MemoCoder              & 17.94\% & 45.52\% & 51.06\% & MemoCoder              & 35.13\%    &    63.59\%   & 74.46\% \\
& MemoCoder w/o Planning   & 17.04\% & 44.89\% &  49.80\% & MemoCoder w/o Planning   & 34.25\%*    &   62.80\%    & 73.70\%\\
& MemoCoder w/o RAG        & 17.65\% & 42.97\%* &  46.53\%* & MemoCoder w/o RAG        & 35.52\%    &  57.59\%*   & 70.83\%*\\
& MemoCoder w/o EP     & 17.81\% & 42.10\%* &  44.69\%* & MemoCoder w/o EP     & 34.77\%    & 59.32\%*    & 68.36\%*\\
\hline
\multirow{3}{*}{MBPP} 
& MemoCoder              & 60.36\%    &  77.69\%   & 83.23\%  & MemoCoder              & 79.56\%    &     89.72\% & 93.70\%\\
& MemoCoder w/o Planning   & 58.70\%*   &  76.19\%   &  81.45\% & MemoCoder w/o Planning  & 79.01\%    &    90.29\%  & 92.88\%\\
& MemoCoder w/o RAG        & 60.62\%    &  74.92\%*  &  77.28\%*& MemoCoder w/o RAG      & 79.40\%    &  87.03\%*   & 92.31\%\\
& MemoCoder w/o EP     & 60.36\%    &  74.17\%*  & 78.42\%* & MemoCoder w/o EP   & 80.21\%    & 87.49\%*    & 91.49\%*\\
\hline
\multirow{3}{*}{HE} 
& MemoCoder              & 33.49\%    &  48.91\%   & 52.40\% & MemoCoder              & 82.97\%    &  92.21\%   & 95.33\% \\
& MemoCoder w/o Planning   & 31.92\%*    &  47.26\%   &  49.33\% & MemoCoder w/o Planning & 81.33\%*   &   91.73\%  & 94.96\%\\
& MemoCoder w/o RAG        & 33.72\%    &  46.44\%*   &  48.10\%* & MemoCoder w/o RAG       & 83.15\%    &  90.48\%*  & 94.13\%\\
& MemoCoder w/o EP     & 33.49\%    &  45.09\%*   &  47.34\%* & MemoCoder w/o EP    & 82.80\%    & 90.79\%*   & 94.47\%\\
\hline
\end{tabular}
\end{table*}

\subsubsection{Results}

\textbf{MemoCoder consistently outperforms the \textit{Zero-Shot} baseline across all datasets}, particularly in the \textit{Pass@10} and \textit{Pass@50} metrics, which reflect the system’s ability to iteratively refine toward correct solutions. For instance, on the \textit{LCB} dataset using LLaMA 3.1-8B, our method achieves a \textit{Pass@10} of \textbf{45.52\%}, compared to \textbf{34.76\%} for Zero-Shot. Similarly, on \textit{MBPP} with Qwen 2.5-32B, our method reaches \textbf{93.70\%} in \textit{Pass@50}, outperforming the Zero-Shot baseline (\textbf{90.12\%}). These gains suggest that MemoCoder is more effective at converging on correct solutions through multiple attempts.

\textbf{In comparison to the \textit{Self-Repair} baseline, which allows repeated attempts but without mentor-driven adaptation or external knowledge, our method also shows consistent improvements.} On \textit{HE} with LLaMA 3.1-8B, for example, \textit{Pass@50} increases from \textbf{47.49\%} (Self-Repair) to \textbf{52.40\%}, while on \textit{LCB} with Qwen 2.5-32B, it improves from \textbf{59.93\%} to \textbf{74.46\%}. Compared to the self-repair baseline, MemoCoder solves problems with fewer iterations and generates fewer hallucinated or redundant error messages. These results suggest that our method is more efficient and produces more stable intermediate outputs during the repair process.

\textbf{Notably, our improvements are less in \textit{Pass@1}, which reflects only the first attempt.} This is expected, as MemoCoder is designed to iteratively refine suboptimal outputs through feedback, not necessarily to produce optimal code in a single step. Thus, while Zero-Shot or Self-Repair may occasionally succeed immediately, our method excels in refining and validating solutions over multiple generations.

\begin{boxA}
    MemoCoder consistently outperforms zero-shot and self-repair baseline methods across all three datasets, particularly in Pass@10 and Pass@50 metrics, demonstrating its strength in iterative code refinement. While Pass@1 results are similar to Self-Repair, MemoCoder shows the self-consistency of MemoCoder. These improvements underscore the value of a multi-agent framework over single-agent zero-shot or self-repair approaches.

\end{boxA}
\subsection{RQ2: What is the impact of each component on the overall performance of MemoCoder?}

\subsubsection{Motivation}

While RQ1 evaluates the overall performance of MemoCoder, it remains important to assess the contribution of each major component. This helps validate our architectural choices and identify which elements are critical for effective code generation and repair. We focus on three configurable modules: the planning mechanism (Planner), error pattern analysis (Mentor), and the retrieval-augmented generation (RAG) module (Mentor). These components represent the core innovations of MemoCoder and can be independently enabled or disabled.

In contrast, components like the Code Writer—responsible for generating and refining code—are fundamental and cannot be ablated. Similarly, low-level utilities such as prompt formatting and error parsing are tightly integrated and not meaningful to isolate. Therefore, we restrict our ablation study to the three modular and conceptually separable components supporting planning, retrieval, and strategy refinement.

\subsubsection{Approach}

To assess the contribution of each component, we conduct an ablation study by disabling one component at a time and evaluating the modified MemoCoder on the \textit{LCB}, \textit{MBPP}, and \textit{HE} benchmarks using pass@k metrics. The three ablated components are:

\begin{itemize}
    \item \textit{Planner:} Produces high-level plans outlining the logical steps to solve a task before code generation.
    \item \textit{Mentor (Error Pattern Analysis):} Abstracts recurring error patterns from past fixes and synthesizes generalized repair strategies for reuse.
    \item \textit{Mentor (Retrieval):} Retrieves relevant past fixes from the knowledge base to guide current error repair using dynamic example-based feedback.
\end{itemize}

\subsubsection{Results}

\textbf{Removing the planning mechanism significantly reduces \textit{Pass@1} in four out of six setups, highlighting its essential role in guiding accurate initial code generation.}  As shown in Table~\ref{tab:rq2_ablation}, this decline is statistically significant ($p < 0.05$) in the following setups: \textit{LCB} and \textit{MBPP} with LLaMA 3.1-8B-Instruct, and \textit{LCB} and \textit{HE} with Qwen 2.5-32B. For example, on \textit{MBPP} with LLaMA 3.1-8B, \textit{Pass@1} drops from 60.36\% to 58.70\%; on \textit{HE} with Qwen 2.5-32B, it falls from 82.97\% to 81.33\%. These results confirm that the planning module helps the model produce more accurate initial outputs. While its absence has limited impact on \textit{Pass@10} and \textit{Pass@50}, likely due to the system’s ability to recover in later iterations, its presence is essential for boosting first-attempt success and could enhance performance even in zero-shot settings.

\textbf{RAG and Error Pattern do not affect Pass@1, but are critical for improving the performance of MemoCoder at Pass@10 and Pass@50.}  
In contrast to the planning component, removing RAG and Error Pattern does not significantly affect Pass@1, but significantly reduces Pass@10 and Pass@50. For instance, when RAG is removed on LCB with Qwen 2.5-32B, Pass@10 drops from 63.59\% to 57.59\%, and Pass@50 falls from 74.46\% to 70.83\%. Similarly, removing the Mentor mechanism reduces Pass@10 from 92.21\% to 90.79\% and Pass@50 from 95.33\% to 94.47\% on HE with Qwen 2.5-32B. 
The influence of RAG and Error Pattern on higher pass rates (\textit{i.e.,} Pass@10 and Pass@50) highlights their role in iterative fixing loops. Since RAG and Error Pattern enhance code repairs across multiple iterations, they significantly improve final solution correctness, particularly in LCB, which contains competition-level questions. The fact that their removal does not affect Pass@1 reinforces the idea that they are not involved in initial code generation but are essential for iterative improvements.


\jack{Do you have the iteration rounds? For example with your knowledge, the rounds should be reduced right?}

\begin{boxA}
Each component of MemoCoder plays a significant and essential role in generating the final solution. Removing the \textit{planning} component significantly reduces Pass@1 in four of six setups: \textit{LCB} and \textit{MBPP} with LLaMA 3.1-8B-Instruct, and \textit{LCB} and \textit{HE} with Qwen 2.5-32B, indicating its importance in producing accurate initial code. Moreover, removing \textit{RAG} or the \textit{Mentor} component significantly impacts code generation performance in Pass@10 and Pass@50, confirming their critical role in iterative error correction. 
\end{boxA}

\subsection{RQ3: How do error types evolve during the code generation process?}
\subsubsection{Motivation}

While the primary objective of MemoCoder is to improve the correctness of generated code, understanding the evolution of error types during the repair process provides deeper insights into how the system handles errors over time. As the code writer agent iteratively refines and re-executes the code, some errors may be successfully resolved, while others persist, evolve, or transform into different types of errors. Analyzing how these error types shift across iterations helps reveal the internal repair dynamics of MemoCoder and enhances its transparency.
Moreover, understanding error transitions allows us to characterize which types of errors are easily handled by the framework and which require further enhancement. 

\jack{I would rephase this as errors, instead of error types. YOu want to study both the types of errors, and the examples in terms of how they evolve. For example, for every 10 or 50 tasks, how many error types or examples do they increase both during the bootstrap or the evaluation phases. This is to hlight the values of your memory, as you still continuously learn but the experience is helping the agent paying back big time}

\subsubsection{Approach}

To investigate the evolution of error types over consecutive iterations, we conduct a qualitative and quantitative analysis of error transitions across repair iterations. For each programming task in our testing datasets, we track the sequence of errors encountered from the initial code generation by the \textit{code writer} agent through each repair attempt by the \textit{code writer} agent. Each time the generated code fails the test cases, we record the iteration number, marking the repair cycle, and 
the type of error (e.g., syntax error, logical error, timeout error)\jack{Is this the same break down as Lei Ma's ICSE work? If yes, just say we use their break down and compare the differences if there are any}. Consequently, from these logs, we generate two key experiments:
\begin{itemize}
    \item \textit{Error evolutions:} We present the distribution of error types across repair iterations over time using a time series plot. This plot reveals which error types are typically resolved early, which persist for longer periods, and whether certain errors dominate the later stages of repair.
    \item \textit{Error transitions:} 
    To analyze how error types evolve over consecutive repair attempts, we construct an error transition matrix. Each row (i.e., y-axis) in the matrix corresponds to a current error type (e.g., \textit{Not Compiled}, \textit{Test Error}), and each column (i.e., x-axis) represents the error type observed in the subsequent iteration. The value at cell $(i, j)$ indicates the proportion of times an error of type $i$ transitions into error type $j$ in the next repair attempt. For example, the cell at row \textit{Not Compiled} and column \textit{Test Error} shows the proportion of times a syntax error was fixed in a way that caused a runtime error in the next iteration. The values are normalized row-wise, meaning each row sums to 1.0 (100\%). A higher value in a cell suggests that errors of a given type are more likely to transform into that specific error category in the next iteration.
    This matrix provides a probabilistic view of error evolution during the repair process and can be interpreted as a first-order Markov chain~\cite{norris1998markov, noei2025empirical}, where each state is an error type and transitions represent changes from one error state to another across repair attempts. Diagonal entries represent cases where the same type of error persists across iterations, while off-diagonal entries capture how errors shift or transform. For instance, a high value in the \textit{Test Failed} to \textit{Pass} cell indicates that many logic errors are resolved within one iteration.\jack{These two figures' scales are wrong, if shown as percentage, it should be in the scale of 0 - 100 instead of 0 to 1}
\end{itemize}
\begin{figure}[h]
    \centering
    \includegraphics[width=\linewidth]{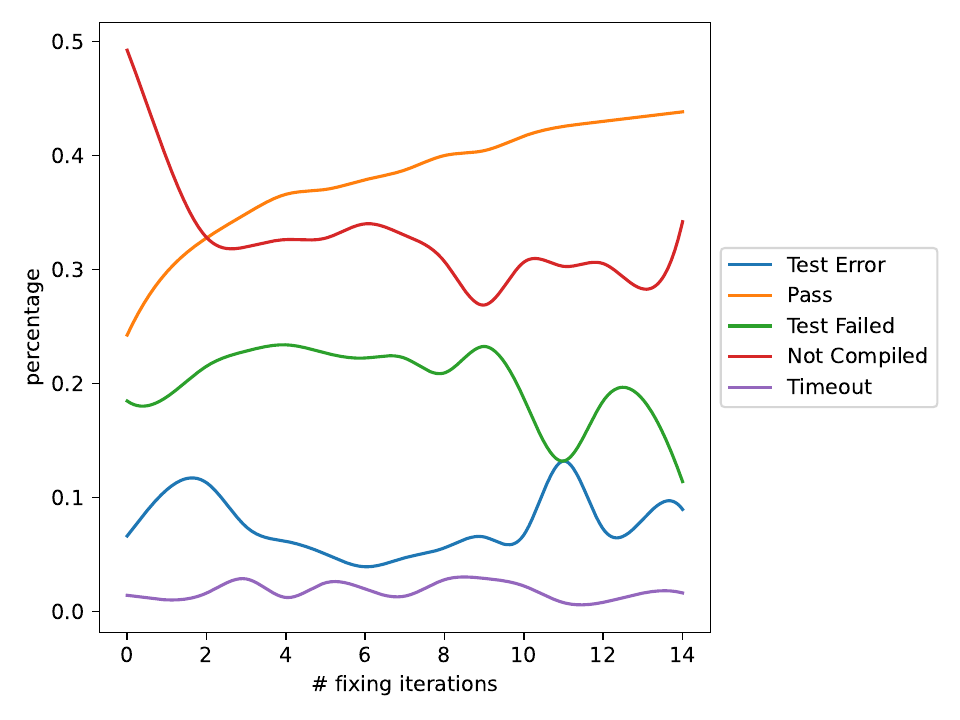}
    \caption{Time series plot of the distribution of errors across repair iterations. The curves are smoothed using a moving average to highlight overall trends.}
    \label{fig:RQ3_line}
\end{figure}
\begin{figure}[h]
    \centering
    \includegraphics[width=\linewidth]{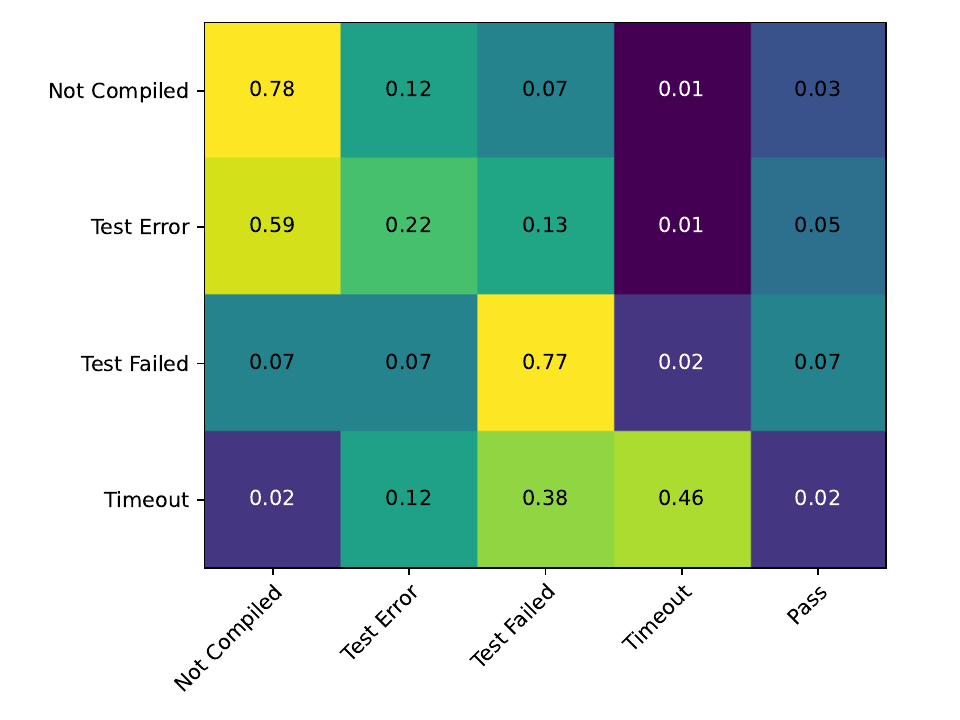}
    \caption{Normalized error type transition matrix. The y-axis represents the current error type, and the x-axis represents the next error state in the subsequent repair iteration. Each cell $(i, j)$ indicates the probability that an error of type $j$ transitions into error type $i$ in the next iteration. }
    \label{fig:RQ3_mat}
\end{figure}

\subsubsection{Results}

\textbf{Over 50\% of the generated code fails to compile in the first round, indicating that many initial outputs contain basic syntax or structural errors.}   Figure~\ref{fig:RQ3_line} shows the time series plot of the distribution of error over continuous iterations.
The percentage of Not Compiled errors drops by 17\% within the first two repair iterations, suggesting that 17\% of the compile errors are simple to fix. Correspondingly, the passing rate increases significantly within the first five repairs.
The trends for other error types further illustrate the system's behaviour during iterative refinement. \textit{Test Failed} errors exhibit a relatively flat pattern over time, suggesting that such errors are more challenging and may require more targeted strategies to resolve. \textit{Test Error} cases (e.g., runtime exceptions) show a modest decline overall, though occasional increases, such as at iteration 11, may reflect the emergence of new issues during the repair process, highlighting the importance of cautious modification. \jack{Can you look into your knowledge DB, what are the \% of breakdown along different error types? Do they have correlation with the resolve rate?}

\textbf{The Not Compiled and Test Failed categories are the most persistent, with over 70\% of cases remaining in the same error type across iterations.}
Figure~\ref{fig:RQ3_mat} shows the probability of transitions between error types (error transition matrix). The error transition matrix provides further insights into how different types of errors evolve.  This persistence indicates that compilation issues and incorrect outputs remain challenging to resolve even after multiple repair attempts. In contrast, \textbf{Test Error and Timeout errors exhibit greater transition rates, frequently changing into other types of errors} rather than persisting in the same category. Therefore, while these errors are easier to modify, they do not always resolve completely and may introduce new issues during the repair process.

\textbf{Test Failed errors have the highest likelihood of being successfully repaired, as 7\% of these cases transition to a passing state in the next round.} This implies that once a function successfully compiles and enters the testing phase, the repair system is more effective at refining the logic to produce the correct output. However, the high persistence of \textbf{Not Compiled} errors,\textbf{78\%} of which remain in the same category in the next iteration, suggests that syntax-related fixes need further improvements, possibly through better leveraging past successful repairs in the Fixing Knowledge set.

\textbf{Timeout errors have a 38\% chance of transitioning to Test Failed errors rather than being directly resolved.} This observation suggests that fixes applied to long-running functions often alter the execution behaviour but do not necessarily produce correct outputs. Improving the repair strategy for handling inefficient or non-terminating functions could further enhance the overall success rate of the system.

\begin{boxA}
In the first round, over 50\% of generations fail to compile, but 17\% of these are resolved within two iterations. \textit{Not Compiled} and \textit{Test Failed} errors persist in 78\% and 77\% of cases, respectively. \textit{Test Error} and \textit{Timeout} show higher transition rates (41\% and 54\%), while 7\% of \textit{Test Failed} cases are successfully repaired in the next iteration.
\end{boxA}

\section{Threats to Validity}
\label{sec:threat}

This section discusses the threats to the validity of our study.

\subsection{Construct Validity}

In MemoCoder, the quality, diversity, and representativeness of the knowledge set directly influence the effectiveness of the repair process. We use over 5,000 training examples from the APPS dataset to construct our initial knowledge set. However, using a larger knowledge set dataset could potentially further improve the performance of MemoCoder.
While we include guiding assertions during generation and repair, this could potentially bias the model toward solving a single example, limiting generalization to the full task specification.

\subsection{Internal Validity}

We evaluate MemoCoder using up to 50 repair attempts (\textit{Pass@50}) and retrieve 10 examples per error type. While alternative limits may affect results, our ablation study (Section~X) shows that performance plateaus beyond 10 retrieved examples and most successful fixes occur within 30 attempts. This supports the reasonableness of our chosen parameters without overfitting.

Another limitation lies in our coarse error taxonomy, which groups failures into broad categories using heuristics. While this design enables scalability and interpretability, it may miss finer-grained distinctions in error behavior. Recent research proposes more detailed taxonomies for LLM-generated errors, which may support better feedback and strategy refinement. Whether a richer taxonomy improves repair quality remains an open question.

\subsection{External Validity}

We use \textit{LLaMA 3.1-8B-Instruct} and \textit{Qwen 2.5-32B} due to their strong performance and relevance. However, evaluating all available LLMs is infeasible, and results may differ with models fine-tuned for specific domains.

Although both models are contamination-free on LCB per the official leaderboard~\cite{lcb}, \textit{MBPP} and \textit{HE} were publicly available prior to these models’ release. Thus, indirect exposure during pretraining is possible, which may influence results on those benchmarks.

We evaluate only Python tasks; effectiveness may vary for other languages such as Java or C++. Our selected benchmarks—\textit{LCB}, \textit{MBPP}, and \textit{HE}—offer diverse tasks but may not reflect the complexity of real-world software development. For example, competitive programming problems often lack long-range dependencies or cross-file logic. In practice, developers work across evolving codebases with richer context and history. MemoCoder has not been tested in such settings, so its generalization to large-scale software maintenance remains an open direction for future research.


\section{Related Work}

\subsection{LLM-Based Code Generation and Its Challenges}

Large Language Models (LLMs) have made substantial progress in code generation, enabling the translation of natural language descriptions into executable code. For example, CodeT5+~\cite{wang2023codet5+} and CodeGen~\cite{nijkamp2022codegen} introduce pretrained encoder-decoder and decoder-only models, respectively, which demonstrate strong performance on standard benchmarks such as HE and MBPP. PanGu-Coder2~\cite{shen2023pangu} extends this line of work with enhanced training objectives and multilingual support.  However, they often struggle with multi-step reasoning and accurate intent alignment, leading to hallucinated outputs~\cite{agarwal2024codemirage,dou2024whats}. Our framework mitigates these issues by introducing structured planning and iterative refinement, improving coherence and logical consistency across generations.

\subsection{Refinement via Fine-Tuning, Prompting, and Iterative Repair}

Refinement techniques include fine-tuning~\cite{pornprasit2024fine}, which adapts LLMs to specific domains, and prompt engineering~\cite{moller2024prompt}, which uses task-specific instructions to steer outputs. Iterative repair approaches like ACR~\cite{zhou2025refinecoder} enhance output quality by repeated critique and correction. However, they typically lack long-term memory. In contrast, our framework incorporates a Fixing Knowledge set and a mentor agent to enable cross-task learning and memory-based refinement.

\subsection{Retrieval-Augmented Generation (RAG)}

Retrieval-Augmented Generation (RAG) enhances LLMs by integrating external knowledge sources into the generation process. By retrieving relevant information from databases or documents, RAG models can produce more contextually accurate and up-to-date code~\cite{guo2023code4uie}. Frameworks like REDCODER have demonstrated the effectiveness of RAG in code generation tasks, particularly in scenarios requiring domain-specific knowledge~\cite{parvez2021redcoder}. However, challenges remain in seamlessly integrating retrieved information and ensuring the relevance and quality of the retrieved data.

\subsection{Multi-Agent Frameworks for Code Generation}

Multi-agent systems leverage the collaboration of specialized agents to tackle complex tasks. Huange \textit{et al.} introduce AgentCoder~\cite{huang2023agentcoder} and MapCoder~\cite{islam2024mapcoder}, which employ multiple agents assigned to distinct roles, such as planning, code generation, testing, and debugging. In their proposed approach, each is responsible for distinct roles such as planning, coding, testing, and debugging. These studies show that this division of labour allows for more robust and efficient code generation processes. Pan \textit{et al.} introduce CodeCoR, which is a self-reflective multi-agent framework that iteratively refines code by evaluating the performance of each agent and their interactions~\cite{pan2025codecor}. Despite the advantages of the multi-agent systems proposed by previous studies, these systems often face challenges in coordinating agents effectively and adapting to new or evolving tasks. Our framework addresses these limitations by integrating a centralized Mentor agent that distills and propagates reusable repair knowledge, coordinates agent interactions through planning and retrieval, and enables adaptive error handling across iterations.

\section{Conclusion}
\label{sec:conclusion}

In this paper, we propose MemoCoder: a memory-augmented multi-agent framework for improving code generation through iterative refinement. MemoCoder incorporates four specialized agents: the \textit{planner}, \textit{code writer}, \textit{test executor}, and \textit{mentor}. 


We evaluate MemoCoder using three benchmark datasets: LCB, MBPP, and HumanEval, and compare its performance against two baseline approaches: zero-shot LLM and a Self-Repair strategy. 
The framework raises \textit{Pass@10} by 12\% and 15\% percentage points and \textit{Pass@50} by 16\% and 19\% points over both baselines, zero-shot LLM and self-repair, respectively, while \textit{Pass@1} is close to Self-Repair. 

We find that removing \textit{planning}, \textit{RAG}, or the \textit{mentor} underscores their importance in refining code correctness over multiple iterations.
Moreover, many syntax errors are resolved within the first few repair attempts. Not Compiled and Test Failed errors remain highly persistent, with over 70\% of cases continuing in the same state. 
Test Error and Timeout cases exhibit higher transition rates, often evolving into different failure types rather than being fully resolved. 

In the future, we aim to extend MemoCoder to support repository-level code generation, where code is synthesized and refined in the context of a full software project.

\bibliographystyle{ACM-Reference-Format}
\bibliography{main}

\end{document}